\documentclass[sigconf]{acmart}

\AtBeginDocument{%
  }

\copyrightyear{2018}
\acmYear{2018}

\usepackage{graphicx}  % Para redimensionamento
\usepackage{adjustbox} % Para ajuste do tamanho da tabela
\usepackage{makecell}  % Para quebras de linha automáticas dentro das células
\usepackage{multirow}
\usepackage{csquotes}
\usepackage[shortlabels]{enumitem}
\usepackage{xcolor}
\usepackage{tcolorbox}
\usepackage{tikz}
\usepackage{longtable}
\usepackage{placeins}
\usepackage{hyperref}

\usetikzlibrary{arrows.meta,positioning}

\tikzset{
  phase/.style={
    draw,
    rounded corners,
    align=center,
    minimum width=4.5cm,
    minimum height=1.3cm,
    font=\small,
    fill=white
  },
  number/.style={
    circle,
    draw,
    fill=gray!60,
    text=white,
    minimum size=8mm,
    inner sep=0pt,
    font=\bfseries
  },
  arrow/.style={
    -{Latex[length=2mm]},
    thick
  }
}

\begin{document}

\title{A Mapping Study About Training in Industry Context in Software Engineering}


\author{Breno Alves de Andrade}
\affiliation{
  \institution{Cesar School}
  \city{Recife}
  \orcid{0009-0007-9364-9317}
  \country{Brazil}
}
\email{baa3@cesar.school}

\author{Rodrigo Siqueira}
\affiliation{
  \institution{cesar.school, Mesa Inc.}
  \city{Recife}
  \orcid{0009-0004-6755-9746}
  \country{Brasil}
}
\email{rms9@cesar.school}

\author{Lidiane C S Gomes}
\affiliation{
  \institution{cesar.school}
  \city{Recife}
  \orcid{0009-0008-9798-956X}
  \country{Brazil}
}
\email{lcsg@cesar.school}

\author{Antonio Oliveira}
\affiliation{
  \institution{cesar.school}
  \city{Recife}
  \orcid{0009-0002-9018-2743}
  \country{Brazil}
}
\email{aaspo@cesar.school}

\author{Danilo Monteiro Ribeiro}
\affiliation{
  \institution{cesar.school, Zup Innovation}
  \city{Recife}
  \orcid{0000-0001-7393-729X}
  \country{Brazil}
}
\email{dmr@cesar.school}

\renewcommand{\shortauthors}{Ribeiro et al.}
%%
%% Article type: Research, Review, Discussion, Invited or position
\acmArticleType{Review}
%%
%% Links to code and data
\acmCodeLink{https://github.com/borisveytsman/acmart}
\acmDataLink{htps://zenodo.org/link}
%%
%% Authors' contribution
\acmContributions{BT and GKMT designed the study; LT, VB, and AP
  conducted the experiments, BR, HC, CP and JS analyzed the results,
  JPK developed analytical predictions, all authors participated in
  writing the manuscript.}
%%
%% Sometimes the addresses are too long to fit on the page.  In this
%% case uncomment the lines below and fill them accodingly.
%%
%% \authorsaddresses{Corresponding author: Ben Trovato,
%% \href{mailto:trovato@corporation.com}{trovato@corporation.com};
%% Institute for Clarity in Documentation, P.O. Box 1212, Dublin,
%% Ohio, USA, 43017-6221}

\begin{abstract}

\textbf{Context:} Corporate training plays a strategic role in the continuous development of professionals in the software engineering industry. However, there is a lack of systematized understanding of how training initiatives are designed, implemented, and evaluated within this domain. \textbf{Objective:} This study aims to map the current state of research on corporate training in software engineering in industry settings, using Eduardo Salas’ training framework as an analytical lens. \textbf{Method:} A systematic mapping study was conducted involving the selection and analysis of 26 primary studies published in the field. Each study was categorized according to Salas’ four key areas: Training Needs Analysis, Antecedent Training Conditions, Training Methods and Instructional Strategies, and Post-Training Conditions. \textbf{Results:} The findings show a predominance of studies focusing on Training Methods and Instructional Strategies. Significant gaps were identified in other areas, particularly regarding Job/Task Analysis and Simulation-based Training and Games. Most studies were experience reports, lacking methodological rigor and longitudinal assessment. \textbf{Conclusions:} The study offers a structured overview of how corporate training is approached in software engineering, revealing underexplored areas and proposing directions for future research. It contributes to both academic and practical communities by highlighting challenges, methodological trends, and opportunities for designing more effective training programs in industry.

\end{abstract}

\maketitle

\section{Introduction}

The acquisition of knowledge, skills, and abilities (KSAs) for present tasks is a key element in enabling individuals to contribute to their organizations and succeed in their current roles \cite{fitzgerald1992training}. 

Research on training consistently highlights two key findings: (a) training is effective, and (b) the design, delivery, and implementation of training programs significantly influence their outcomes \cite{salas2012science}. In contemporary organizations, where rapid innovation is crucial, continuous learning and skill development have become important. Companies invest in training because they recognize that a qualified and constantly evolving workforce is a strategic advantage \cite{salas2012science}.

This need is particularly pressing in the Information Technology (IT) sector. IT firms face growing pressure to build and sustain high-quality human capital, which is essential for maintaining innovation and competitiveness \cite{diniz2024skill}. However, a persistent mismatch between the offerings of educational institutions and the demands of the software industry has led to a global shortage of adequately trained professionals \cite{diniz2024skill}. This gap forces companies to invest heavily in internal training programs to close the skill deficit \cite{diniz2024skill}.

Human capital is one of the most strategic assets for any software organization. Decisions about how to develop this capital—whether by upskilling existing staff, hiring new talent, outsourcing tasks, or adopting a hybrid strategy—are central to sustaining performance \cite{boudreau2005talentship, wingreen2005assessing}. Nonetheless, despite the strategic importance of training, education and workforce development are often not core concerns in software engineering companies \cite{wingreen2005assessing}.

Although numerous studies discuss corporate training initiatives in the software engineering industry, the current body of knowledge remains fragmented. There is no consolidated effort to systematize the area as a whole, nor to provide a broad overview that allows researchers and practitioners to understand its structure, trends, and knowledge gaps. This fragmentation can lead to several consequences: redundant or isolated efforts, low transferability of findings, limited strategic alignment with organizational goals, and missed opportunities to inform policy and theory-building.

To address this gap, this study presents a systematic mapping study that investigates how training has been conducted, evaluated, and reported within the software engineering industry. Using Salas’ framework—which organizes training activities into four categories: \textit{Training Needs Analysis}, \textit{Antecedent Training Conditions}, \textit{Training Methods and Instructional Strategies}, and \textit{Post-Training Conditions}—we analyzed 26 primary studies to identify trends, gaps, and opportunities.

This study contributes by offering an organized perspective on industrial training in software engineering and by highlighting underexplored areas that may benefit from future investigation. It provides researchers and practitioners with actionable insights to design training programs that are better aligned with organizational strategy and professional development goals.

The remainder of this paper is structured as follows. Section~\ref{section:background} provides the background and related work. Section~\ref{section:method} describes the methodology and procedures used in this study. Section~\ref{section:results} presents the results, including answers to the research questions. Section~\ref{section:discussion} discusses the findings, their implications, and Section~\ref{section:limitations} the limitations. Finally, Section~\ref{section:conclusion} concludes the paper and suggests areas for future research.

\section{Background}\label{section:background}

\subsection{Definitions}
The terms \textbf{training}, \textbf{education}, \textbf{development}, and \textbf{learning} are frequently used interchangeably, yet each has a distinct meaning within the human resources domain. Clarifying these concepts is essential for understanding their role in corporate training and workforce development.

\textbf{Training} refers to practical, task-oriented instruction aimed at improving performance in current roles. It is typically hands-on and focuses on the transmission of knowledge, skills, and abilities (KSAs) required for job-related tasks \cite{masadeh2012training}. According to Fitzgerald, training is “(a) the acquisition of knowledge and skill for present tasks, (b) a tool to help individuals contribute to the organization and be successful in their current positions, and (c) a means to an end” \cite{fitzgerald1992training}. Effective training must be linked to performance outcomes and result in behavioral changes observable in the workplace \cite{fitzgerald1992training}.

\textbf{Education}, in contrast, is conceptually broader and more reflective. It emphasizes understanding, critical thinking, and theoretical knowledge rather than direct application. Education often takes place in formal academic settings and is influenced by lifelong social interactions that help individuals adapt to dominant norms and values \cite{masadeh2012training, dos2014educaccao}.

\textbf{Development} encompasses both training and education and adopts a long-term perspective. It aims to prepare individuals for future roles by enhancing their KSAs through a continuous process involving learning, reflection, and practical experience \cite{fitzgerald1992training}. Development initiatives often align with career progression strategies and organizational sustainability goals.

Despite their differences, training, education, and development share a central concern with learning \cite{garavan1997training}. According to the Cambridge Dictionary, \textbf{learning} is the acquisition of knowledge through study or experience and is characterized by behavioral change resulting from external stimuli \cite{garavan1997training}. Clear distinctions among these terms support more effective human resource strategies and organizational planning \cite{masadeh2012training}.

In this context, \textbf{competencies} refer to the integrated set of KSAs required for effective job performance. Parry defines competencies using three foundational components \cite{parry1996quest}:
\textbf{Knowledge}: Conceptual and procedural understanding — the \textit{know-how}.\textbf{Skills}: Practical capabilities developed through experience — the \textit{can-do}.
\textbf{Abilities}: Innate or developed traits enabling performance — the \textit{capable of doing}.

Understanding education and its underlying purpose requires analyzing the broader sociocultural and organizational context in which educational practices and discourses are situated \cite{eboli2004educaccao}.

\textbf{Corporate education} has emerged as a strategic response to the growing mismatch between higher education outcomes and industry demands. It addresses the rapid obsolescence of knowledge and the need for continuous workforce qualification \cite{masalimova2014multi}. It is defined as an internal training system aligned with organizational goals, structured to support managers and specialists at all levels \cite{masalimova2014multi}.

Organizations increasingly invest in training as a means of achieving competitive advantage. Empirical evidence shows that both on-the-job and off-the-job training initiatives significantly improve individual performance and bridge critical skill gaps \cite{}. Structured progams suces job enrichment initiatives have been shown to enhance adaptability and competence when aligned with business objectives. These benefits, alongside the need to mitigate knowledge obsolescence, explain the growing adoption of corporate education systems \cite{masalimova2014multi, shafiq2017effect}.

In summary, training and development activities benefit both employees—by improving career trajectories—and organizations—by increasing operational efficiency and resilience \cite{shafiq2017effect}.

Because training is a systemic process involving pre-training, delivery, and post-training phases, this study adopts Salas' framework to organize the literature review. This model offers a structured approach for categorizing training practices and outcomes, helping to align analysis with the realities of the software engineering industry \cite{petersen2008systematic}.

\subsection{Salas' Framework}\label{subsection:salas-framework}
The framework proposed by Salas offers a comprehensive perspective on organizational training, drawing from extensive research conducted from the early 1990s to the 2000s. It emphasizes advances in training theory, instructional methods, and the critical role of post-training activities. The model highlights the influence of cognitive, motivational, and organizational factors, as well as the growing integration of technological tools in training delivery \cite{salas2001science}.

Table~\ref{tab:salas_areas} provides a brief overview of the four main components of Salas' framework.

\begin{table}
    \caption{Salas' Framework}
    \label{tab:salas_areas}
    \centering
    \renewcommand{\arraystretch}{1.5}
    \setlength{\tabcolsep}{7pt}
    \begin{adjustbox}{max width=0.45\textwidth}
        \begin{tabular}{|c|m{7cm}|} 
            \hline 
            \textbf{Category} & \textbf{Description} \\
            \hline
            \makecell[c]{Training Needs \\ Analysis} & This first step in training development focuses on the process of deciding who and what should be trained. \\
            \hline
            \makecell[c]{Antecedent Training \\ Conditions} & Individual factors before training, such as cognitive ability, self-efficacy, and goal orientation, have been particularly influential in shaping behavior and outcomes. \\
            \hline
            \makecell[c]{Training Methods and \\ Instructional Strategies} & It incorporate a combination of tools, methods, and content that together form a structured instructional approach. \\
            \hline
            \makecell[c]{Post-Training \\ Conditions} & Events that occur to evaluate training, and on examining the events that ensure transfer and application of newly acquired knowledge, skills, and abilities (KSAs). \\
            \hline
        \end{tabular}
    \end{adjustbox}
\end{table}

\subsubsection{Training Needs Analysis}\label{subsubsection:training-need-analysis} 
The first phase of training development involves identifying the essential elements required for successful intervention. This includes determining who needs training, what should be taught, and how the training aligns with organizational goals \cite{salas2012science}. Training Needs Analysis (TNA) plays a crucial role in optimizing resource allocation and maximizing training effectiveness.

\paragraph{Organizational Analysis}\label{paragraph:organization-analysis}
This dimension focuses on the alignment between training goals and broader organizational factors such as strategic objectives, available resources, constraints, and support for learning transfer \cite{salas2001science}.

\paragraph{Job/Task Analysis}\label{paragraph:job-task-analysis}
This analysis identifies the specific KSAs required to perform particular tasks effectively, considering the expected outcomes and role-specific demands \cite{salas2001science}.

\subsubsection{Antecedent Training Conditions}\label{subsubsection:antecedent-training-conditions}
Antecedent conditions refer to the factors present before training that influence its success. These include individual characteristics and contextual variables that shape learners’ readiness and engagement. Such conditions are as critical as the training itself in determining outcomes \cite{salas2001science}.

\paragraph{Individual Characteristics}\label{paragraph:individual-characteristics}
Trainees bring to the learning environment distinct traits such as cognitive ability, self-efficacy, and goal orientation—all of which significantly influence training engagement and learning outcomes \cite{salas2001science}.

\paragraph{Training Motivation}\label{paragraph:training-motivation}
Motivation to learn is shaped by both individual variables (e.g., age, personality) and organizational factors (e.g., climate, support). Higher motivation leads to greater knowledge retention and behavioral change \cite{salas2012science}.

\paragraph{Training Induction and Pretraining Environment}\label{paragraph:training-induction-pretraining}
Pre-training strategies, such as advanced organizers, structured preparation, and cognitive scaffolding, are essential to optimize learning conditions and ensure that trainees are equipped to absorb and apply new information effectively \cite{salas2001science}.

\subsubsection{Training Methods and Instructional Strategies}\label{subsubsection:training-methods}
Instructional strategies encompass the design and application of tools, content, and delivery techniques to improve learning retention and job performance.

\paragraph{Specific Learning Approaches}\label{paragraph:specific-learning-approaches}
These include pedagogical techniques such as feedback loops, distributed practice, and reinforcement, which are used to increase the retention and transfer of learning.

\paragraph{Learning Technologies and Distance Training}\label{paragraph:learning-technologies}
Training approaches increasingly leverage digital technologies—such as video conferencing, e-learning platforms, and online courseware—to enable flexible and scalable delivery \cite{salas2001science}.

\paragraph{Simulation-Based Training and Games}\label{paragraph:simulation-training}
Simulations replicate real-world environments and tasks, offering immersive learning experiences that foster deep engagement, reduce error rates, and improve performance \cite{salas2001science}.

\paragraph{Team Training}\label{paragraph:team-training}
Team-based training focuses on building collaborative skills, fostering communication, and enhancing group performance. It prepares individuals to apply learned skills in various team contexts, improving both individual and collective effectiveness.

\subsubsection{Post-Training}\label{subsubsection:post-training}
Post-training conditions are vital to ensuring that acquired KSAs are retained and applied in the workplace. The long-term impact of training is strongly linked to the follow-up and evaluation processes \cite{salas2001science}.

\paragraph{Training Evaluation}\label{paragraph:training-evaluation}
Evaluation involves measuring the effectiveness of training through behavioral, cognitive, and affective indicators. Well-defined objectives and assessment criteria ensure alignment with performance goals \cite{salas2012science}.

\paragraph{Transfer of Training}\label{paragraph:transfer-of-training}
Training transfer refers to the application, generalization, and maintenance of KSAs in the work context. It is a core objective of any training initiative and directly impacts organizational effectiveness \cite{salas2001science}.

\subsection{Training in the Software Engineering Industry}

Technology companies face increasing pressure to develop and sustain high-quality human capital in order to maintain innovation and competitiveness \cite{diniz2024skill}. Factors such as software quality, delivery speed, and cost efficiency have become central to market differentiation \cite{ruhe1999experience}. In this context, continuous training is a strategic necessity, requiring dedicated structures aligned with ongoing organizational learning efforts.

\textbf{Software engineering} as a discipline exhibits characteristics that distinguish it from other engineering fields, particularly in the context of workforce development:

\begin{enumerate}
    \item \textbf{Software engineering is an experimental discipline}: Knowledge must be continuously refined through real-world development experiences, which are then translated into educational and training programs \cite{ruhe1999experience}.
    \item \textbf{Software is developed, not manufactured}: Success depends heavily on human expertise and creativity, making education and training essential for adopting new technologies \cite{ruhe1999experience}.
    \item \textbf{Each software product is unique}: Training programs must be contextualized to reflect the specific challenges, goals, and technological environments of the organization \cite{ruhe1999experience}.
\end{enumerate}

The integration of training, education, development, and learning is therefore particularly relevant in software engineering. The human-centered, evolving, and context-specific nature of software development demands tailored strategies for capacity building. As organizations increasingly recognize the link between human capital strategies \cite{diniz2024skill} and product outcomes \cite{ruhe1999experience}, aligning training approaches with strategic goals becomes critical for fostering innovation and sustainable performance \cite{fitzgerald1992training}.

For the purposes of this study, we adopt the definition of \textbf{corporate education} as a long-term organizational strategy focused on the continuous development of KSAs through a structured set of training initiatives aligned with business goals. In this perspective, training addresses operational competencies (\textit{what to do}), training programs define tactical approaches (\textit{how to do it}), and corporate education sets strategic direction (\textit{where to go}). Together, they form an integrated framework essential for building and sustaining a competitive workforce.
demonstrate

\subsection{Related Work}
This underscores the ongoing
challenge of translating training into workplace performance—an
issue similarly relevant in software engineering, where collaboration and coordination are critical, yet the impact of training remains
insufficiently studied.
Some research highlights the importance of training in the software industry. The relationship between professional training, human capital development, and the challenges of the software industry has been addressed from different perspectives over the past years. Arora and Athreye\cite{ARORA2002253} highlight the importance of the software industry in fostering human capital and increasing productivity within the Indian economy. According to the authors, the sector not only contributes to job creation and exports but also drives innovative organizational models and private investments in training, directly influencing corporate governance and professional qualification.

Kulkarni et al.\cite{Kulkarni2010} provide an empirical perspective on the initial training offered by Indian technology companies to recent graduates. The study reveals a misalignment between traditional academic education and industry expectations, which leads companies to invest heavily in their own training programs. In addition to technical skills, corporate training programs also address behavioral competencies and cultural aspects relevant to working in distributed teams.

More recently, Chethana’s review \cite{chethana2023review} analyzed training structures using diverse sources of gray literature—including Google Scholar publications, symposium materials, financial reports, official websites, news portals, and social media testimonials—focusing on training programs in Indian companies from 2012 to 2021. Although methodologically distinct from our systematic approach, their proposed framework aligns closely with Salas’ foundational training model \cite{salas2001science}, comprising:

\begin{enumerate}
    \item Needs assessment for training and development,
    \item Clear objective setting,
    \item Methodology selection and detailed planning,
    \item Program implementation,
    \item Results analysis, and
    \item Ongoing monitoring.
\end{enumerate}

Salas et al. (2006) \cite{salas2006does} reviewed 28 studies on Crew Resource Management (CRM), a structured training model originally developed in aviation and later adopted in other high-risk domains such as medicine, offshore oil, and nuclear power. CRM focuses on enhancing non-technical skills—including communication, teamwork, and decision-making—to improve safety and performance. Using Kirkpatrick’s evaluation framework \cite{kirkpatrick1970evaluation}, Salas et al. found that while trainees generally responded positively to CRM training, the results regarding learning outcomes, behavioral changes, and organizational impact were mixed

These related works form our "golden set"—serving both as foundational references for constructing our search strings and as primary sources for snowballing procedures.

\section{Method}\label{section:method}
In this section, we present an adapted and applied \textbf{systematic mapping study} in software engineering, based on the guideline.\cite{petersen2008systematic}. The study was carried out in five main steps: 
(1) Definition of Research Questions, (2) conducting the Search, (3) Screening of Papers, (4) Keywording using Abstracts, (5) Data Extraction and Mapping Process.

\subsection{Research Questions}
The primary goal of this systematic mapping study is to provide an overview of what the cientific literature are talking about the Training in Industry Context in Software
Engineering and identify  research and results available of the approches and impacts of training of competency in the professional for software engineering research and results available. Additionally, the study aims to identify with areas of Salas' framework the article are located, the main challenges of the area map publication frequencies over time to observe trends and identify the forums where research in the area has been published. These objectives are reflected in the following research questions, which guide this study:

\begin{itemize}
    \item \textbf{RQ1} - What evidence does the scientific literature provide about of training for software engineering professionals in industry contexts?
    \begin{itemize}
        \item \textbf{RQ1.1}- How  the articles were categorized according to the research areas defined in Salas' framework? (\ref{subsection:RQ1.1})
        \item \textbf{RQ1.2} - Which variables are the authors analyzing for the development of the training?(\ref{subsection:RQ1.2})
        \item \textbf{RQ1.3} - What are the objectives of the trainings? (\ref{subsection:RQ1.3})
    \end{itemize}
\end{itemize}

\textbf{RQ1} is the primary goal of this systematic mapping study is to identify evidence-based practices and critical gaps in corporate training for software engineering professionals. This overarching question guides our investigation of training approaches, their impacts, and implementation challenges.

\textbf{RQ1.1} aims to provide an overview by categorizing the articles according to their research areas. This helps to clarify the distribution of studies, highlighting both well-explored and underrepresented areas within the literature.

\textbf{RQ1.2} focuses on identifying the variables analyzed by the authors during the development of training. Understanding these variables is essential to grasp how training programs are structured and what factors are considered critical for their success.

\textbf{RQ1.3} investigates the main objectives of the training initiatives described in the literature. This allows us to understand the intended outcomes of such programs and how they align with industry needs.

\subsection{Conduct Search for Primary Studies}
Our primary studies were identified using search strings in automatic databases or by manually searching \cite{petersen2008systematic} and also snowballing our references. Our research was conducted using four authoritative software engineering digital libraries: ACM and IEEE. We selected them because ACM is widely used in the field of computer science,  IEEE is a leading applied engineering repository, Scopus is the largest multidisciplinary index, and Springer is a high-impact publisher—ensuring comprehensive coverage of both technical and theoretical perspectives. 

Our automated search strategy targets three core domains: training, software engineering, and industry applications. To ensure comprehensive coverage,
we applied the OR operator to include synonyms of the key terms
and the AND operator to ensure that the concepts and context
were combined effectively.

The search string was designed to query only the \textbf{title} and \textbf{abstract} fields in two digital library services: IEEE Xplore Digital Library and ACM Digital Library.

We used the following synonyms to build the query for \textbf{training}:  
corporate training, corporate education, on-the-job training, in-house training, job training, professional training, training in companies, employee education, professional development, training program, educational program, IT training, personnel development, institutional education, bootcamp, corporate university, and development of human capital.

For \textbf{software engineering}, we included:  
software development, IT, and software engineering.  

For \textbf{industry}, we considered:  industry, IT firms, IT industry, corporate, organizations, and workplace.

This query is designed to include topics related to corporate education, software engineering, and industry. These terms provide a comprehensive coverage of key areas such as workforce training, software development, and industrial applications, making sure that the search results are relevant and aligned with these important fields.

The process was conducted in four main stages: \textbf{Automated Search Strategy}, \textbf{Primary Study Selection}, \textbf{Full Read} and \textbf{Snowballing}. The automated search yielded 574 articles from ACM and 1,571 from IEEE, totaling 2,145 articles. In all selection stages, two researchers analyzed the articles individually and compared the results at the end. A third researcher, with more experience in systematic reviews and education in Software Engineering, resolved conflicts regarding the selected articles, ensuring a general agreement between both researchers. In the first stage, the selection was based on the title and abstract. In the next stage, the analysis included the previous fields complemented by the conclusion, reducing the number to 126 articles selected.

\begin{figure}
    \centering
    \begin{adjustbox}{max width=0.45\textwidth}
        \begin{tikzpicture}[node distance=1cm and .8cm]

            \node[phase] (p1) {\textbf{Automated Search Strategy} \\ ACM $n=574$\\ IEEE $n=1571$ \\ $n=\textbf{2145}$ };
            \node[phase, below=of p1] (p2) {\textbf{Primary Selection}\\ title and abstract \\ $n=\textbf{126}$};
            \node[phase, right=of p1] (p3) {\textbf{Full Read} \\ full body content \\ $n=\textbf{17}$};
            \node[phase, right=of p2] (p4) {\textbf{Snowballing}\\ title and abstract \\ $n =\textbf{29}$};;
            \node[phase, right=of p3] (p5) {\textbf{Snowbolling Full Read}\\ full body content \\ $n = \textbf{9}$};;
            \node[phase, right=of p4] (p6) {\textbf{Results}\\ total of papers \\ $n =\textbf{26}$};

            % Numbers
            \node[number, anchor=south east, xshift=.2cm, yshift=-.3cm] at (p1.north west) (n1) {1};
            \node[number, anchor=south east, xshift=.2cm, yshift=-.3cm] at (p2.north west) (n1) {2};
            \node[number, anchor=south east, xshift=.2cm, yshift=-.3cm] at (p3.north west) (n1) {3};
            \node[number, anchor=south east, xshift=.2cm, yshift=-.3cm] at (p4.north west) (n1) {4};
            \node[number, anchor=south east, xshift=.2cm, yshift=-.3cm] at (p5.north west) (n1) {5};
            \node[number, anchor=south east, xshift=.2cm, yshift=-.3cm] at (p6.north west) (n1) {6};

            % Arrows
            \draw[arrow] (p1) -- (p2);
            \draw[arrow] (p2) -- (p3);
            \draw[arrow] (p3) -- (p4);
            \draw[arrow] (p4) -- (p5);
            \draw[arrow] (p5) -- (p6);
        \end{tikzpicture}
    \end{adjustbox}
    \caption{Phases and study counts}
\end{figure}
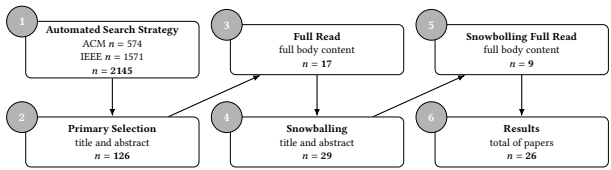

The process was conducted in four main stages: \textbf{Automated Search Strategy}, \textbf{Primary Study Selection}, \textbf{Full-Text Analysis}, and \textbf{Snowballing}. The automated search yielded 574 articles from ACM and 1,571 from IEEE, totaling 2,145 articles. 

In all selection stages, two researchers independently analyzed the articles and compared results, with a third experienced researcher resolving conflicts to ensure consensus. The first stage filtered articles based on titles and abstracts. The second stage added conclusions to the evaluation criteria, reducing the set to 126 articles. 

In the third phase, we performed a full-text reading and selected 12 articles that met all the criteria. In the fourth phase, the snowballing technique was applied, resulting in the selection of 5 additional relevant articles from the 12 analyzed. Additionally, 4 more articles were included from our references, totaling 9 articles obtained through snowballing.
  
After completing all stages, the systematic mapping resulted in \textbf{26 studies} (12 from full-text analysis + 9 from snowballing)

\begin{figure}
    \centering
    \includegraphics[width=1\linewidth]{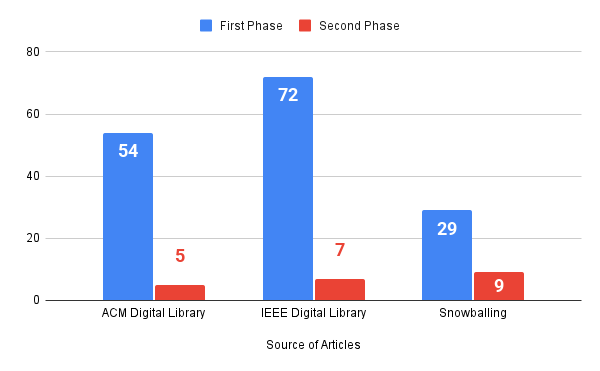}
    \caption{Conduct Search}
    \label{fig:enter-label}
\end{figure}

\subsection{Screening of Papers for Inclusion and Exclusion}
Inclusion and exclusion criteria were used to exclude studies that were not relevant to answering the research questions \cite{petersen2008systematic}. 

\textbf{Inclusion Criteria:}
\begin{itemize}
    \item Primary studies from the software engineering industry focusing on corporate education, which propose analyses, frameworks, methods, training approaches, or evaluations.
\end{itemize}

\textbf{Exclusion Criteria:}
\begin{itemize}
    \item Articles not published in peer-reviewed venues.
    \item Articles in languages other than English or Portuguese.\footnote{Portuguese is the mother language of the researchers.}
    \item Articles related to IT support.
    \item Articles not containing relevant terms in the title and abstract.
    \item Articles not directly related to corporate education applied in the software engineering industry.
    \item Articles that were not implemented or applied in practice.
    \item Articles conducted exclusively with academic students.
    \item Articles involving both students and professionals in which it is not possible to isolate the results related to professionals
    \item Articles that address the IT field in a generic way, alongside other areas.
\end{itemize}
The inclusion and exclusion criteria were defined to ensure the relevance and the quality of the selected sample. Only studies from the software engineering industry with an explicit focus on training were included, as the main goal of this research is to understand how training is designed, implemented, and evaluated within the specific context of the software engineering industry.  These methodological decisions aimed to ensure that the final set of studies was representative, coherent with the research objectives.

\subsection{Data Extraction and Mapping of Studies}
In this step, we used a Google Sheets table to document the data extraction process. The table contained two tabs with identical column fields, where each researcher entered the data for the papers. A third tab, called \textbf{Compilation}, was created by combining the data from both researchers, including resolved conflicts and consensus reached with the advisor acting as a mediator.

When the reviewers entered the data of a paper into the scheme, they provided a short rationale and evidence explaining why the paper should be classified under a certain category of Salas' areas.

A data extraction form was created to extract the necessary information to answer the research questions.

\section{Results}\label{section:results}

This section presents the findings of the systematic mapping study. \textbf{A total of 26 primary studies} focused on corporate education in software engineering—specifically those conducted and implemented by industry—were selected and analyzed. The complete list of selected articles and their identifiers is presented in Table~\ref{tab:selected-works}.

\begin{figure}
    \centering
    \includegraphics[width=1\linewidth]{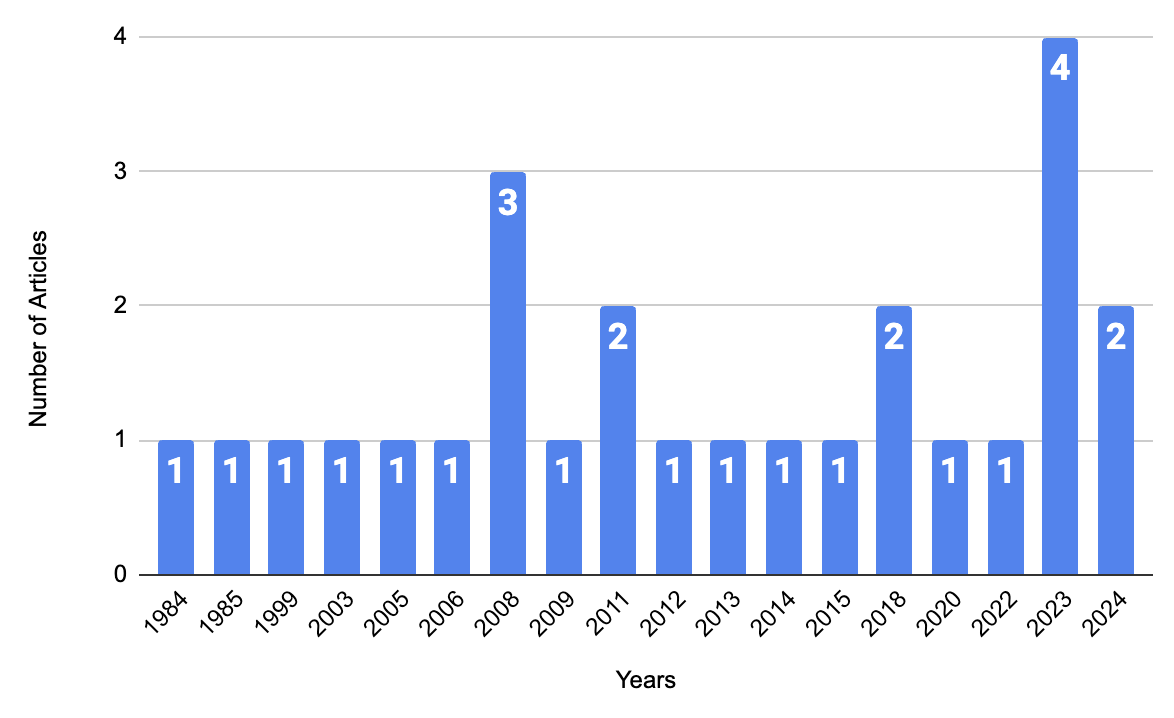}
    \caption{Distribution of articles by year of publication}
    \label{fig:articles-per-year}
\end{figure}

To identify temporal trends in this research domain, the selected articles were also organized by year of publication, as illustrated in Figure~\ref{fig:articles-per-year}. The earliest study dates back to \textbf{1984}. A modest rise in publication volume were observed in the years \textbf{2008, 2011, 2018, 2023 and 2024}, while the other years showed low or no activity in this topic area. Despite its relevance to industry, the topic has received little systematic investigation in the academic literature.

\begin{figure}
    \centering
    \includegraphics[width=1\linewidth,alt={Distribution of Training Research Across Salas' Framework Categories} ]{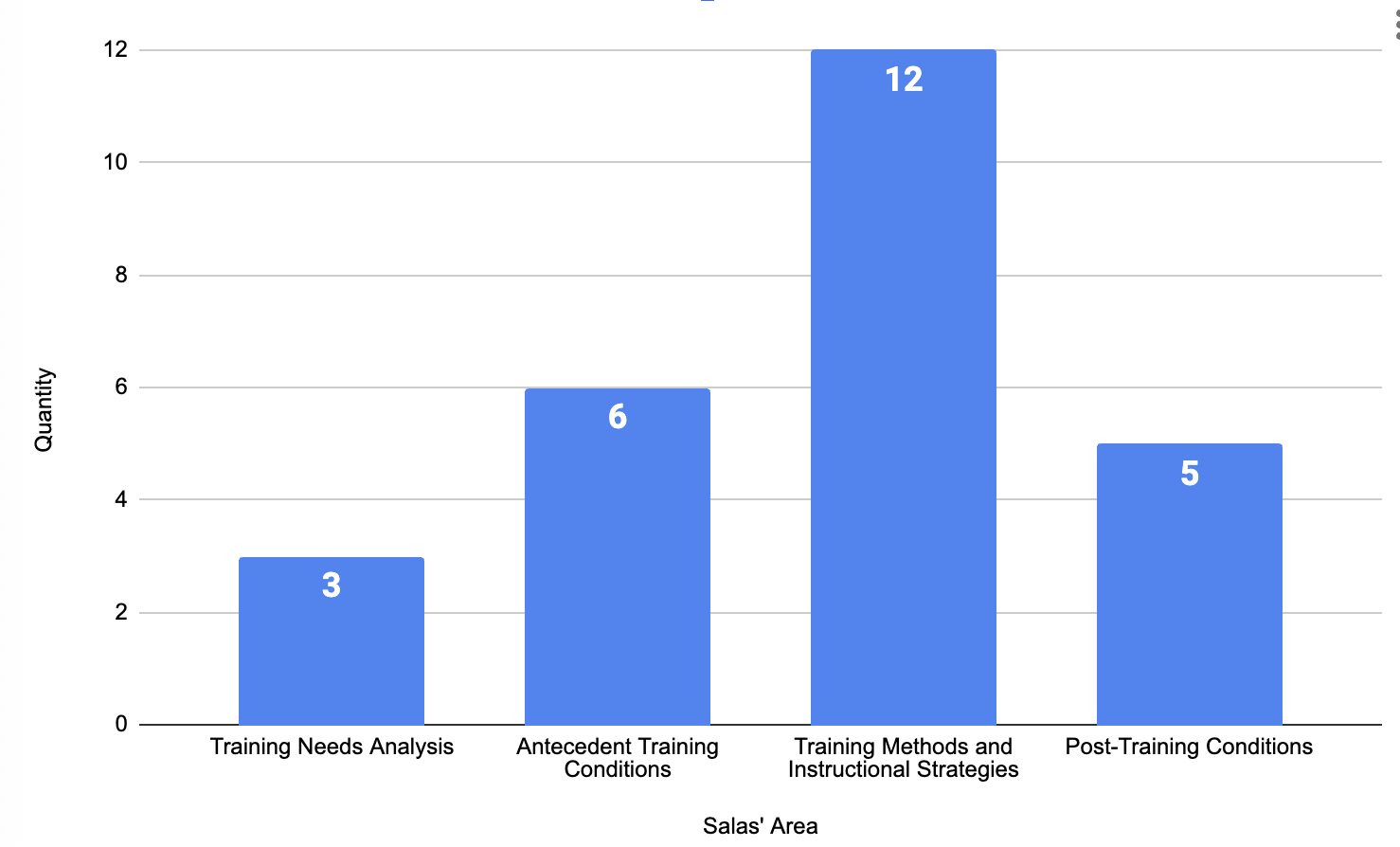}
    \caption{Distribution of Training Research Across Salas’ Framework}
    \label{fig:distribution}
\end{figure}

To structure the analysis, each study was classified according to the categories defined in Salas’ framework for corporate training: \textbf{Training Needs Analysis}, \textbf{Antecedent Training Conditions}, \textbf{Training Methods and Instructional Strategies}, and \textbf{Post-Training Conditions}. The distribution of studies these categories is shown in figure~\ref{fig:distribution}.

\renewcommand{\arraystretch}{1.3} % um pouco mais compacto ainda
\setlength{\tabcolsep}{3pt} % menos espaço horizontal entre colunas

\enlargethispage{-1cm} % Reduz a altura da página atual para "forçar" o encaixe da tabela

\begin{table}
\centering
\caption{Selected articles and their identifiers}
\label{tab:selected-works}
\renewcommand{\arraystretch}{1.3} % Ajusta altura das linhas
\setlength{\tabcolsep}{4pt} % Ajusta espaço entre colunas
\begin{adjustbox}{max width=\linewidth}
\begin{tabular}{|c|p{12cm}|}
\hline
\textbf{Ref.} & \textbf{Title} \\
\hline
\cite{boyeena2011blended} & A blended approach to course design and pedagogy to impart soft skills: An IT company's experiences from software engineering course \\
\hline
\cite{wingreen2006} & A Q-Methodological Study of IT Professionals’ Person-Organization Fit for Training and Development \\
\hline
\cite{mason2003} & Aligning workforce development software process improvement strategy for accelerated adoption of software engineering capability \\
\hline
\cite{tsukamoto2011analysis} & Analysis of the Motivation of Learners in the In-House Training of Programming in Japanese ICT Industries \\

\hline
\cite{wingreen2005} & Assessing the IT Training and Development Climate: An Application of the Q-methodology \\
\hline
\cite{silva2023corporative} & Corporative Education Used to Attract and Qualify Workforce for the Software Industry \\
\hline 
\cite{hudepohl2014deploying} & Deploying an online software engineering education program in a globally distributed organization \\ 
\hline 
\cite{wolfschwenger2022design} & Design and Evaluation of an Agile Framework for Continuous Education in Software Engineering \\
\hline 
\cite{schaul1985design} & Design using software engineering principles: overview of an educational program \\
\hline 
\cite{agarwal2008employees} & Employees Perception towards Training in IT Sector \\
\hline
\cite{jan2012employees} & Employees' training and development in IT sector: An essential instrument for effectiveness of organizational productivity \\
\hline
\cite{ruhe1999experience} & Experience Factory-based professional education and training \\
\hline
\cite{bondi2009experience} & Experience with Training a Remotely Located Performance Test Team in a Quasi-agile Global Environment \\
\hline 
\cite{bapna2013human} & Human Capital Investments and Employee Performance: An Analysis of IT Services Industry \\ 
\hline 
\cite{prikladnicki2008improving} & Improving Contextual Skills in Global Software Engineering: A Corporate Training Experience \\ 
\hline 
\cite{wolfschwenger2023integrating} & Integrating Cloud-Based AI in Software Engineers' Professional Training and Development \\ 
\hline 
\cite{wingreen2018professionals} & IT professionals' person–organization fit with IT training and development priorities \\ 
\hline
\cite{goelzer2024nurturing} & Nurturing Talent: The IT Academy Journey into Quality Development \\ 
\hline
\cite{stanojeska2024personalized} & Personalizes training for success implementation of advanced technologies in the IT industry \\ 
\hline
\cite{ulema1984planning} & Planning for in-house software engineering education \\ 
\hline 
\cite{zubereva2023} & Professional Development of IT Industry Specialists at the Workplace: Trends, Focus and Prospects \\
\hline
\cite{morales2018requirements} & Requirements Analysis Skills: How to Train Practitioners? \\
\hline
\cite{sinha2020role} & Role of leadership in enhancing the effectiveness of training practices: Case of Indian information technology sector organizations \\
\hline
\cite{samuel2015synchronous} & Synchronous training in distributed software development team \\ 
\hline 
\cite{berenbach2008evaluation} & The Evaluation of a Requirements Engineering Training Program at Siemens \\
\hline

\cite{ribeiro2023understanding} & Understanding Self-Efficacy in Software Engineering Industry: An Interview Study \\ 
\hline 

\end{tabular}
\end{adjustbox}
\end{table}

\subsection{RQ1.1 - How  the articles were categorized according to the research areas defined in Salas' framework?}\label{subsection:RQ1.1}

To classify the selected studies within Salas’ framework, each training initiative was analyzed according to the specific subareas that comprise the four main categories. This classification was based on the evidence reported in each article—such as objectives, methods, and implementation strategies—indicating alignment with one or more subareas. Table~\ref{tab:areas-subareas-articles} presents the mapping between the reviewed studies and the corresponding areas of Salas’ framework.

\begin{table}
    \centering
    \caption{Mapping between Salas' Areas and Articles}
    \label{tab:areas-subareas-articles}
    \renewcommand{\arraystretch}{1.3}
    \setlength{\tabcolsep}{4pt}
    \begin{adjustbox}{max width=.45\textwidth}
        \begin{tabular}{|
            >{\centering\arraybackslash}m{3cm} |
            >{\centering\arraybackslash}c |
            >{\centering\arraybackslash}m{2cm} |}
            \hline
            \textbf{Area} & \textbf{Sub-Area} & \textbf{Articles} \\
            \hline
            \multirow{2}{*}{\makecell{Training Needs \\ Analysis}} 
            & Organizational Analysis 
            & \cite{ruhe1999experience, prikladnicki2008improving, ulema1984planning} \\
            \cline{2-3}
            & Job/Task Analysis & - \\
            \hline
            \multirow{3}{*}{\makecell{Antecedent Training \\ Conditions}} 
            & Individual Characteristics & \cite{wingreen2018professionals,  wingreen2006} \\
            \cline{2-3}
            & Training Motivation & \cite{tsukamoto2011analysis} \\ 
            \cline{2-3}
            & \makecell{Training Induction and \\ Pretraining Environment} & \cite{sinha2020role, stanojeska2024personalized,wingreen2005} \\ 
            \hline 
            \multirow{4}{*}{\makecell{Training Methods and \\ Instructional Strategies}} 
            & Specific Learning Approaches 
            & \cite{schaul1985design,wolfschwenger2022design,mason2003, wolfschwenger2023integrating,morales2018requirements,goelzer2024nurturing} \\
            \cline{2-3}
            & \makecell{Learning Technologies and \\ Distance Training} 
            & \cite{hudepohl2014deploying, bondi2009experience, samuel2015synchronous, zubereva2023} \\ 
            \cline{2-3}
            & \makecell{Simulation-Based \\ Training and Games} 
            & - \\ 
            \cline{2-3}
            & Team Training & \cite{boyeena2011blended, silva2023corporative} \\ 
            \hline 
            \multirow{2}{*}{\makecell{Post-Training \\ Conditions}} 
            & Training Evaluation 
            & \cite{bapna2013human, ribeiro2023understanding, jan2012employees, agarwal2008employees} \\
            \cline{2-3}
            & \makecell{Transfer of Training} & \cite{berenbach2008evaluation} \\
            \hline 
        \end{tabular}
    \end{adjustbox}
\end{table}

\subsubsection{\textbf{Training Needs Analysis}}

Three studies were classified under the subarea \textbf{Organizational Analysis}, as they emphasize strategic planning to either optimize existing training practices or design new programs aligned with organizational needs:

\begin{itemize}
  \item \cite{ruhe1999experience} applies an organizational analysis approach using the Goal-Question-Metrics (GQM) method to assess the effectiveness of training programs.
  \item \cite{prikladnicki2008improving} proposes a strategic plan for implementing training initiatives that address the global needs of a software company.
  \item \cite{ulema1984planning} presents a corporate education plan focused on software engineering, highlighting the need to adapt training content to organizational contexts.
\end{itemize}

\subsubsection{\textbf{Antecedent Training Conditions}}
We identified works in three subareas: \textbf{Individual Characteristics}, \textbf{Training Motivation} and \textbf{Training Inductions and Pretraining Environment}.
In the \textbf{Individual Characteristics}.
\begin{itemize}
  \item \cite{wingreen2018professionals} investigates behavioral and perceptual differences among IT professionals, particularly focusing on the concept of Person–Organization (P–O) fit. The study argues that even when professionals possess similar technical skills, experience, and compensation, their perception of alignment with organizational values may vary significantly. These perceptions play a central role in shaping training motivation and engagement, underscoring the importance of individual characteristics in training outcomes.

    \item \cite{wingreen2006} The seven person–organization (P–O) fit perspectives revealed by this study demonstrate that IT professionals hold personal beliefs about their alignment with IT training and development programs that span multiple dimensions — across areas such as content, venue, and resources — rather than representing a unitary and one-dimensional concept.
  
\end{itemize}

In the \textbf{Training Motivation}:
\begin{itemize}
  \item\cite{tsukamoto2011analysis} investigates with the motivational attitude was compared across different groups based on criteria such as gender, academic background, and prior programming experience.
\end{itemize}

In the \textbf{Training Inductions and Pretraining Environment}:
\begin{itemize}
  \item \cite{sinha2020role} investigates how leadership styles — transformational and transactional — influence the effectiveness of training practices in the IT sector in India. 
  \item\cite{stanojeska2024personalized} shows that personalized training programs significantly increase employee engagement and the effectiveness of new technology implementation, strengthening organizational performance and competitiveness.
   \item \cite{wingreen2005} investigates with the Q-methodology to develop a tool for the assessment of the IT training and development climate, identify and interpret natural groups of IT professionals according to their own professional development priorities to be used as a baseline for managerial decision making, and make recommendations for its use as a managerial decision tool.

\end{itemize}

\subsubsection{\textbf{Training Methods and Instructional Strategies}}

The selected articles were classified into three subareas: \textbf{Specific Learning Approaches}, \textbf{Learning Technologies and Distance Training}, and \textbf{Team Training}. No articles were associated with the \textbf{Simulation-Based Training and Games} subarea.

For \textbf{Specific Learning Approaches}, a variety of strategies were identified:

\begin{itemize}
    \item \cite{wolfschwenger2022design} employed Problem-Based Learning (PBL) to promote learning through real-world problem-solving activities related to DevOps practices.
    \item \cite{schaul1985design} described a workshop model focused on applying mathematical models for functions and data abstractions in software programs.
    \item \cite{wolfschwenger2023integrating} adopted a hands-on approach using Artificial Intelligence to deliver personalized feedback, interactivity, and adaptability to individual learning styles.
    \item \cite{morales2018requirements} implemented a mentoring-based training plan in Requirements Engineering to help professionals acquire practical knowledge in the field.
    \item \cite{goelzer2024nurturing} emphasized hands-on learning within software quality training, integrating practices like BDD and agile methodologies.
    \item \cite{mason2003} utilized a class combination to training professionals in software engineering offering flexibility distance training.
\end{itemize}

In the \textbf{Learning Technologies and Distance Training} subarea, the following approaches were observed:

\begin{itemize}
    \item \cite{hudepohl2014deploying} utilized video content, e-learning platforms, and webinars to train globally distributed software teams.
    \item \cite{bondi2009experience} conducted remote training sessions for a performance testing team located across different regions.
    \item \cite{samuel2015synchronous} explored both synchronous and asynchronous training formats for distributed teams, highlighting how communication strategies supported effective knowledge acquisition.
    \item \cite{zubereva2023} explored Smart educational environment involves the integration of innovative concepts, intelligent hardware and software, smart classrooms equipped with the latest technologies, and educational processes based on modern and intelligent teaching and learning strategies.
    
\end{itemize}

In the \textbf{Team Training} subarea:

\begin{itemize}
    \item \cite{boyeena2011blended} developed group-based training programs in which participants were evaluated on both technical and soft skills.
    \item \cite{silva2023corporative} provided training as part of a recruitment and retention process for new employees in software engineering roles.
\end{itemize}

\subsubsection{\textbf{Post-Training Conditions}}

This area includes two subareas: \textbf{Training Evaluation} and \textbf{Transfer of Training}.

In The \textbf{Training Evaluation}, focusing on individual development and organizational outcomes:

\begin{itemize}
    \item \cite{bapna2013human} demonstrated that corporate training programs can improve individual performance.
    \item \cite{agarwal2008employees} The study shows that there are seven major factors which
affect the perception of employees towards training in the
IT sector. These factors include Training Duration,
Training Method, Training Impact, Training Outcome,
Job Performance, Training benefit and Training
Effectiveness.
    \item \cite{jan2012employees} The research through Structural
Equation Modeling (SEM) using AMOS infers that the good training programmes will increase the
productivity of the organization.
    \item \cite{ribeiro2023understanding} found that bootcamp-style training improved team self-efficacy, enhanced deliverable quality, reduced lead times, and encouraged knowledge sharing.
\end{itemize}

The \textbf{Transfer of Training} subarea was addressed by \cite{berenbach2008evaluation}, which analyzed the practical application of training content in the participants' daily work activities.

\subsection{RQ1.2 - Which variables are the authors analyzing for the development of the training?}\label{subsection:RQ1.2}

Three main variables were identified across the studies: training duration, subjects covered, and methods or tools used to support the training.

\begin{itemize}
    \item \textbf{Training Duration}: The length of the training programs varied significantly across studies. For example, \cite{berenbach2008evaluation} reported a short-term course of three days, \cite{boyeena2011blended} presented a five-day program, \cite{schaul1985design} described a two-week training, and \cite{silva2023corporative} reported a six-month program. This variation may reflect differences in pedagogical goals, participant profiles, and organizational contexts.

    \item \textbf{Subjects Covered}: The content of the training programs often focused on core software engineering areas. Requirements Engineering appeared in \cite{boyeena2011blended}, \cite{hudepohl2014deploying}, and \cite{berenbach2008evaluation}, while Software Testing was featured in \cite{boyeena2011blended}, \cite{hudepohl2014deploying}, and \cite{bondi2009experience}.

    \item \textbf{Methods and Tools}: Various tools and approaches were reported to support training. In the \textbf{Training Needs Analysis} category, \cite{ruhe1999experience} used the Goal-Question-Metrics (GQM) method, and \cite{prikladnicki2008improving, ulema1984planning} employed structured interviews to capture stakeholder expectations. In the \textbf{Antecedent Training Conditions} category, \cite{wingreen2018professionals, wingreen2006} used Q Methodology to explore person-organization fit. In the \textbf{Training Methods} category, workshops \cite{schaul1985design}, PBL \cite{wolfschwenger2022design}, hands-on activities \cite{wolfschwenger2023integrating, goelzer2024nurturing, bondi2009experience}, and e-learning platforms \cite{boyeena2011blended, hudepohl2014deploying, samuel2015synchronous, morales2018requirements, silva2023corporative} were frequently used. For \textbf{Post-Training Evaluation}, \cite{berenbach2008evaluation} employed the Kirkpatrick Model, \cite{bapna2013human} used the PCMM method, and \cite{ribeiro2023understanding} conducted interviews.
\end{itemize}

\subsection{What are the objectives of the trainings?}\label{subsection:RQ1.3}

The analyzed papers reveal study groups with common themes, although directed toward different goals faced by the industry. Below are the objectives of the studies reviewed.

\begin{itemize}
    \item {\textbf{Software Engineering Practices}: The paper \cite{boyeena2011blended} aims to align corporate training with organizational demands, focusing on the technical and behavioral development of new talents, based on the 5 main areas of the SWEBOK.}
    \item {\textbf{Distributed Training}: The article \cite{hudepohl2014deploying} address to implement an effective online education program for software engineers in a globally distributed organization. The article \cite{bondi2009experience} focuses on conducting a testing training program with two distributed teams—one located in India and the other in New Jersey, United States. The main challenges of this training were cultural differences and time zone discrepancies. The article \cite{prikladnicki2008improving} reports on a training experience carried out in a multinational company to develop the technical team's skills and highlights the same challenges mentioned in \cite{bondi2009experience}. Finally, the article \cite{samuel2015synchronous} presents the use of synchronous and asynchronous training formats to support geographically distributed teams.} 
    \item {\textbf{DevOps and Cloud}: The article \cite{wolfschwenger2022design} focuses to apply an agile, hands-on framework to promote the continuous education of software engineers, with a focus on DevOps and cloud computing.}
    \item {\textbf{ Use of Artificial Intelligence}: The article \cite{wolfschwenger2023integrating} aims to apply generative AI to enhance productivity and quality during training for software developers.}
    \item {\textbf{Requirements Engineering}: The article \cite{morales2018requirements} goals to develop a training plan for educating professionals in the discipline of Requirements Engineering.}
    \item {\textbf{Software Quality}: The article \cite{goelzer2024nurturing} objectives to develop a quality mindset in professionals participating in the training, emphasizing the implementation of quality best practices from the early stages of the software development process. In turn, the article \cite{schaul1985design} focuses on the need to introduce software engineering principles and the application of mathematical models in the design of training programs to reduce defects and costs, while improving software quality.}
\end{itemize}

\section{Discussion}\label{section:discussion}
\subsection{General}
The analysis reveals that studies in the software engineering industry address all four categories proposed by Eduardo Salas as essential for corporate training. However, not all subareas within each category are equally represented. This uneven distribution highlights both the maturity of certain research directions—such as \textbf{Training Methods and Instructional Strategies}—and the absence of empirical work in other foundational subareas.

Specifically, in the \textbf{Training Needs Analysis} category, only the subarea \textit{Organizational Analysis} was addressed \cite{ruhe1999experience, prikladnicki2008improving, ulema1984planning}, while \textit{Job/Task Analysis} was not explored in any of the selected studies. The lack of focus on job/task-level needs may compromise the relevance and precision of training initiatives, since identifying specific competencies required for job performance is a key starting point for effective program design.

In the \textbf{Antecedent Training Conditions}, it was observed that some subareas were addressed by only a few studies. Individual Characteristics were explored in two articles \cite{wingreen2018professionals,  wingreen2006}, while Training Motivation was examined in only two studies \cite{wingreen2005,tsukamoto2011}. The subarea Pretraining Environment and Induction was covered in two other studies \cite{sinha2020role, stanojeska2024personalized}. Despite the occasional presence of these investigations, the scarcity of research in these subareas is concerning, as these factors directly influence learner readiness, engagement, and learning effectiveness. Overlooking such conditions may result in training programs that are misaligned with the participants’ actual needs, thereby compromising knowledge transfer and the intended organizational impact.

For \textbf{Training Methods and Instructional Strategies}, all subareas except \textit{Simulation-Based Training and Games} were addressed. The use of simulations can provide realistic and controlled learning environments that are especially useful for complex, high-stakes contexts—yet no study in the review employed or evaluated this method. This omission suggests an opportunity for future work to investigate the applicability and impact of simulation-based training in software engineering contexts.

In the \textbf{Post-Training Conditions} category, both subareas were covered, but to varying extents. \textit{Training Evaluation} was addressed by three studies \cite{bapna2013human, ribeiro2023understanding, jan2012employees, agarwal2008employees}, while \textit{Transfer of Training} appeared in only one \cite{berenbach2008evaluation}. This limited attention to transfer raises concerns, as the real value of training lies in its application to daily work. Without examining how new knowledge and skills are transferred to the job context, it is difficult to assess whether training achieves its intended impact.

Compared to other domains—such as healthcare, manufacturing, and Industry 4.0—where corporate training has been more extensively studied and systematized \cite{cazeri2022training}, the field of software engineering still lacks a cohesive body of literature on this topic. This highlights the novelty and importance of the present study in offering a first structured overview based on an established training framework.

Moreover, the scarcity of longitudinal assessments is notable. Most reviewed studies emphasize immediate outcomes rather than long-term effects on performance, behavior, or retention. This absence reinforces the need for future research to adopt longitudinal designs that can better capture training impact over time.

The lack of studies exploring training aligned with individual learning needs and the reliance on anecdotal evidence also suggest that the academic field is still in an early stage of development. It may be premature to speak of a consolidated research tradition on corporate training in software engineering. Therefore, future studies might benefit from adapting or extending existing frameworks—such as Salas’—to build context-specific models that reflect the nuances of software development environments.

A possible explanation for the limited number of studies found in this mapping is the lack of publication of empirical research by industry. Companies often develop and implement internal training programs but do not share these initiatives with the academic community. This behavior may be driven by factors such as strategic information protection, market competitiveness, or even the absence of incentives or partnerships with research institutions. As a result of this lack of publication, there is little contribution to the consolidation of scientific knowledge on training in Software Engineering. This limitation directly impacts the ability to generalize findings and hinders the advancement of the field, which lacks more robust empirical studies connected to real-world industry practices. Therefore, it is essential to strengthen the connection between academia and industry, fostering collaborations that enable the documentation and critical analysis of training practices that are already in use but remain invisible in the scientific literature.

Furthermore, since companies rarely disclose training outcomes publicly, an alternative approach could involve extracting relevant data from annual and financial reports to obtain more primary-level insights - mirroring Chethana's successful approach of analyzing alternative sources \cite{chethana2023review} through diverse gray literature. This suggests that exploring supplementary data channels may offer valuable pathways to bridge the current research gap in corporate training studies.

Another relevant point observed in the selected studies is the absence of reports on negative experiences or failed training initiatives. This lack of critical reflection limits the field’s maturity, as it prevents researchers and practitioners from learning from challenges, design flaws, or contextual mismatches that might undermine training effectiveness. Including such cases in future studies would contribute to a more comprehensive understanding of what works, what doesn’t, and under which conditions.

This mapping offers researchers an updated view of the state of training initiatives in the software engineering industry. By identifying underexplored areas and patterns of instructional design, this study also serves as a reference for professionals responsible for designing training programs. It encourages evidence-based decision-making and highlights theoretical foundations that can inform and enhance future corporate training efforts.

\subsection{Implications for Industry, Society, and Academia}

\textbf{For Industry}: This study presents a practical framework that organizations can use to evaluate and improve their corporate training strategies. By highlighting underexplored areas—such as job/task analysis and Simulation-Based Training and Games can better align training programs with organizational goals and employee needs. Additionally, the findings point to the importance of structured evaluation methods and long-term follow-up to ensure return on investment. 

\textbf{For Society}: The quality of software development directly affects the reliability and safety of digital systems used across all sectors of society. Improving training for software engineers ultimately supports the creation of better digital services, enhances cybersecurity practices, and contributes to economic development. Furthermore, by encouraging inclusive and context-aware training practices, this research promotes the development of more equitable and accessible digital solutions.

\textbf{For Academia}: This work identifies critical research gaps and proposes directions for future studies. It lays the groundwork for a research agenda in corporate training for software engineering, advocating for more rigorous methodologies, longitudinal designs, and interdisciplinary approaches. Scholars can build on this mapping to explore the interaction between human factors, organizational behavior, and training outcomes, as well as to develop and validate training frameworks tailored to the software industry context.

\section{Threats to Validity and Limitations}\label{section:limitations}

As with any systematic mapping study, this research is subject to a number of limitations and potential threats to validity. To address these, we followed guidance from relevant methodological literature \cite{wohlin2014guidelines, zhou2016validity}.

\textbf{Search and Selection Bias}: There is always a risk that relevant studies were missed due to the use of specific keywords or databases. To mitigate this, we built the search string based on Salas' framework and existing reviews, and conducted both automatic and manual searches.

\textbf{Data Extraction Bias}: The process of classifying studies and extracting evidence may be influenced by the researchers' interpretations. To reduce this threat, data extraction was performed in pairs, with a third researcher consulted in cases of disagreement.

\textbf{Publication Bias}: Our study only included published research, which may favor positive findings and omit null or negative results. This is a known issue in systematic reviews \cite{zhou2016validity}, and should be considered when interpreting the frequency and distribution of findings.

\textbf{Generalizability}: This mapping focused on the field of software engineering. Although the use of Salas' framework provides a generalizable lens, findings may not reflect practices in other industries. Moreover, the use of Salas' framework outside its original organizational psychology context introduces potential limitations in fit or interpretation.

\textbf{Lack of Methodological Rigor in Primary Studies}: Many of the included papers are experience reports rather than empirical evaluations. As such, the quality and reliability of the evidence varies, which limits the strength of inferences about effectiveness or impact.

Future studies should seek to replicate this mapping in other domains, employ quality assessment checklists during selection, and explore cross-sector comparisons to strengthen the field’s theoretical and practical base.
\section{Conclusion}\label{section:conclusion}

This study mapped the existing literature on corporate training in software engineering applied to industry, organizing the selected works according to Eduardo Salas’ training framework. Through the analysis of 26 articles, the study identified prevailing themes, emerging patterns, and critical gaps in the development and evaluation of training programs targeting software engineering professionals.

The findings reveal a predominant focus on the category of \textbf{Training Methods and Instructional Strategies}, with a notable diversity in pedagogical approaches such as e-learning, hands-on activities, workshops, and problem-based learning. However, significant gaps remain in other categories—particularly \textbf{Training Needs Analysis}, \textbf{Antecedent Training Conditions}, and \textbf{Post-Training Conditions}—highlighting the lack of studies on areas such as \textit{Job/Tasks Analysis} and \textit{Simulation-Based Training and Games}

The study also observed that although some methods were employed to assess training needs, individual profiles, instructional delivery, and outcomes, there is insufficient evidence on the effectiveness and consistency of these methods across different organizational contexts. Most contributions take the form of experience reports, limiting generalizability and empirical robustness.

Additionally, authors considered practical aspects of training design—such as duration, structure, and curriculum content—but there is little standardization among the programs. This heterogeneity reflects the contextual nature of corporate education in the field but also underscores the need for more systematic and comparative research.

In doing so, this mapping contributes to the systematization of knowledge in the field by offering a structured overview of current practices and uncovering opportunities for future investigation. It serves as a resource for both researchers and practitioners aiming to design, implement, and evaluate more effective training initiatives within software engineering environments.

As a next step, we recommend expanding the scope of the review to include additional databases, gray literature, and potentially non-English publications. Future studies may also benefit from incorporating quality assessment criteria, conducting meta-analyses where applicable, and examining training initiatives through longitudinal or experimental designs to better understand their long-term impact on professional performance and organizational outcomes.

This research was originally developed in Portuguese. The authors used OpenAI's ChatGPT and DeepSeek Chat exclusively as support tools for translating the content into English and improving the cohesion and clarity of the text.

\section*{Artefacts Disponibles}

 Sheet with the extracted datas: 
\href{https://docs.google.com/spreadsheets/d/1EnBi0Y1OgTHX4zyBWXOo1jclAwKPtIZ9JHqo5W6cYjc/edit?usp=sharing}{\textbf{Google Sheets Link}}.

\bibliographystyle{ACM-Reference-Format}
\bibliography{references}

\end{document}